\documentclass[a4paper,11pt]{article}
\usepackage{latexsym}
\usepackage{cite}
\usepackage{epsfig}
\usepackage{float}
\usepackage{amssymb}
\author{H. Mohseni Sadjadi \footnote{mohsenisad@ut.ac.ir}
\\ {\small Department of Physics, University of Tehran,}
\\ {\small P. O. B. 14395-547, Tehran 14399-55961, Iran}}
\title {On cosmic acceleration in four-dimensional Einstein-Gauss-Bonnet gravity}
\begin{document}
\maketitle
\begin{abstract}
We study the possibility that in the model introduced in \cite{GBN}, the Gauss-Bonnet term alone gives rise to the cosmic acceleration and super-acceleration in four-dimensional FLRW space-time at the late time. We also discuss transitions from deceleration to acceleration and acceleration to super-acceleration. We show that the Gauss-Bonnet invariant can drive the acceleration in the low redshift provided that its
regularized coefficient has the same order as the squared of the ratio of the reduced Planck mass to the Hubble parameter.
\end{abstract}
\section{Introduction}

It has been more than two decades that we have found that the expansion of the Universe is positively accelerating \cite{acc1,acc2}. Many models have been proposed to describe this acceleration in line with astrophysical data that has become much more accurate and complete in recent years. The negative pressure required for this expansion cannot be described in the framework of the standard cosmology with the known particles in the standard model of particle physics. Therefore some authors introduced exotic matter such as scalar field\cite{quin1,quin2,quin3,quin4,quin5}, and so on, in the framework of Einstein standard model of gravity, while some others modified the usual gravitational model \cite{mod1,mod2,mod3,mod4,mod5}. Meanwhile, many authors used both methods simultaneously to correct each model's defects according to observational data\cite{cb1,cb2,cb3,cb4}. So in the literature, we encounter actions comprising many complicated scalar terms made by combinations of matter fields and (modified) geometrical functions of Riemann curvature, Einstein tensor, Ricci curvature, torsion, etc. One of these geometrical terms is the Gauss-Bonnet term. As the variation of this term with respect to the metric vanishes in four dimensions, it does not modify the Einstein equation and does not alter the dynamics of the system. That is why, instead of using the pure Gauss-Bonnet term, which is a total derivative, modified Gauss-Bonnet model \cite{mGB1,mGB2,mGB3,mGB4} or generalized models in which the Gauss-Bonnet term is coupled to other fields have been used to study the cosmic acceleration \cite{mGBc1,mGBc2,mGBc3,mGBc4,mGBc5,mGBc6}.

Recently a new model has been proposed in which the Gauss-Bonnet term appears with a factor $1/(D-4)$ in the Lagrangian, where $D$ is the dimension of the space-time \cite{GBN}. As the contribution of Gauss-Bonnet in the equations of motion is proportional to $D-4$, inserting this factor produces a nontrivial contribution in the Einstein equations in four dimensions. In other words, the infinity is eliminated by a zero of the same order in the classical equation of motion. This elimination looks somehow similar to the dimensional renormalization technique used in quantum field theory, where quantum infinite are canceled out by adding appropriate counter-terms to the Lagrangian. However, in dimensional regularization, the dimension is treated as a continuous parameter. Despite the singular coefficients in the Lagrangian, it is the equations of motion that describe the classical behavior of the system. In the last few months, there have been many discussions and debates about [1]'s proposal, while many subjects have been studied in this framework such as black hole solutions and their stability \cite{GBNB0,GBNB1,GBNB2,GBNB3,GBNB4,GBNB5,GBNB6,GBNB7,GBNB8,GBNB9,GBNB10,GBNB11}, quasi-normal modes \cite{GBNQ},  wormhole solutions \cite{GBNW1,GBNW2}, cosmological evolution \cite{GBNC1,GBNC2,GBNC3,GBNC4}, stellar structures \cite{sss}, dust collapse \cite{dc}, holographic phase transition \cite{hol}, etc.

In this paper, we will consider the possible acceleration of the Universe in the context of this novel model. In this study, we employ only the Gauss-Bonnet term and do not involve other dark energy candidates or exotic fields, to see how this term alone can lead to an acceleration in the cosmic evolution. The scheme of the paper is as follows:
In the second section, we first introduce the model and present a detailed computation to derive the Friedmann equations in Friedmann-Lemaitre-Robertson-Walker (FLWR) space-time. Then by using these equations, we study the dynamics and acceleration of the Universe, the critical points, and their stabilities. By studying the deceleration parameter, we investigate possible deceleration to acceleration and acceleration to super-acceleration transitions.

We use units $\hbar=c=1$ through the paper.

\section{Friedmann equations in four-dimensional EGB model}
We consider the Einstein-Gauss-Bonnet (EGB) action
\begin{equation}\label{1}
S=\int d^Dx\sqrt{-g}\left(\frac{M_P^2R}{2}\right)+S_{GB}+S_m
\end{equation}
$M_P$ is the reduced Planck mass which in terms of the gravitational constant is $M_P=\sqrt{\frac{1}{8\pi G}}$. Note that in $D$ dimensions the mass dimension of $M_P$ is $[M_P]=D/2-1$,  and  for $D=4$ we have  $M_P=2.436\times 10^{18}GeV$.  $S_{GB}$ is the Gauss-Bonnet action
\begin{equation}\label{3}
S_{GB}=\int d^D x \sqrt{-g} \alpha \mathcal{G}
\end{equation}
in which $\alpha$ is a constant and $\mathcal{G}$ is the Gauss-Bonnet invariant term
\begin{equation}\label{4}
\mathcal{G}=R^{\mu \nu}_{\rho \sigma}R^{\rho \sigma}_{\mu \nu}-4R^{\mu}_{\nu}R^{\nu}_{\mu}+R^2.
\end{equation}
The mass dimension of $\alpha$ is $[\alpha]=D-4$. In $D=4$, $\alpha$ is dimensionless. In our study $\alpha$ is a constant although in extended models it may be a function of dynamical fields. We have placed all matter components (baryonic and dark) in $S_m$.

By variation of the action with respect to the metric, we obtain
\begin{equation}\label{ref1}
R_{\mu \nu}-\frac{1}{2}g_{\mu \nu}R=\frac{1}{M_P^2}T^{m}_{\mu \nu}+\frac{2}{M_P^2}\frac{1}{\sqrt{-g}}\frac{\delta S_{GB}}{\delta g^{\mu \nu}}
\end{equation}
whose the trace is given by \cite{GBN}
\begin{equation}\label{ref2}
(1-\frac{D}{2})R=\frac{1}{M_P^2}\left(T^m+\alpha (D-4)\mathcal{G}\right)
\end{equation}
where $T^m$ is the trace of matter energy momentum tensor $T^m_{\mu \nu}$. In a four-dimensional space-time, and for a singular $\alpha$, such that $\alpha (D-4)$ is still finite, the Gauss-Bonnet term participates in field equations.

Variation of the Gauss-Bonnet action $S_{GB}$ with respect to the metric gives
\begin{equation}\label{5}
\frac{g_{\nu \rho}}{\sqrt{-g}}\frac{\delta S_{GB}}{\delta g_{\mu \rho}}=\mathcal{A}^\mu_\nu+\frac{1}{2}\mathcal{G}\delta^\mu_\nu
\end{equation}
where $\mathcal{A}^\mu_\nu$ is
\begin{equation}\label{6}
\mathcal{A}^\mu_\nu=-2{R^{\mu\alpha}}_{\rho \sigma}{R^{\rho \sigma}}_{\nu \alpha}+4{R^{\mu \alpha}}_{\nu \beta}{R^{\beta}}_{\alpha}+4{R^{\mu}}_{\alpha} {R^{\alpha}}_{\nu}-2R{{R^\mu}}_\nu
\end{equation}

To determine (\ref{5}) we need to compute the Riemann curvature tensor, the Ricci tensor, the Ricci scalar, and the Gauss-Bonnet term.
In a D-dimensional spatially flat FLRW space-time
\begin{equation}\label{2}
ds^2=-dt^2+a^2(t)(dx_1^2+dx_2^2+dx_3^2+.....+dx_{D-1}^2),
\end{equation}
where $a(t)$ is the scale factor, by computing the Riemann curvature tensor components, we find
\begin{equation}\label{7}
R_{i0i0}=-a\ddot{a},\,\,\,  R_{ijij}=a^2\ddot{a}^2
\end{equation}
Leading to following components for the Ricci curvature
\begin{equation}\label{8}
R_{00}=-(D-1)\frac{\ddot{a}}{a},\,\,\,\  R_{ii}=(D-2)\dot{a}^2+a\ddot{a}
\end{equation}
So the scalar curvature is obtained as
\begin{equation}\label{9}
R=(D-1)(D-2)\frac{\dot{a}^2}{a^2}+2(D-1)\frac{\ddot{a}}{a}
\end{equation}
The Gauss-Bonnet scalar, is derived as
\begin{eqnarray}\label{10}
\mathcal{G}&=&R^{\mu \nu}_{\rho \sigma}R^{\rho \sigma}_{\mu \nu}-4R^{\mu}_{\nu}R^{\nu}_{\mu}+R^2\nonumber \\
&=&(D-3)(D-2)(D-1)\left(\left(D-4\right)\frac{\dot{a}^4}{a^4}+\frac{4\dot{a}^2\ddot{a}}{a^3}\right)
\end{eqnarray}
Using (\ref{7}), (\ref{8}), and (\ref{9}), the tensor ${\mathcal{A}^\mu}_\nu$, is computed as
\begin{eqnarray}\label{11}
{\mathcal{A}^i}_i&=&-2(D-2)(D-3)\left((D-4)\frac{\dot{a}^4}{a^4}+3\frac{\dot{a}^2\ddot{a}}{a^3}\right)\nonumber \\
{\mathcal{A}^0}_0&=&-2(D-1)(D-2)(D-3)\frac{\dot{a}^2\ddot{a}}{a^2}
\end{eqnarray}
So by using (\ref{10}) and (\ref{11}) the contribution of the Gauss-Bonnet term to the equations of motion is given by
\begin{eqnarray}\label{12}
\frac{g_{i \rho}}{\sqrt{-g}}\frac{\delta S_{GB}}{\delta g_{i \rho}}&=&\frac{1}{2}(D-2)(D-3)(D-4)\left((D
-5)\frac{\dot{a}^4}{a^4}+\frac{4\dot{a}^2\ddot{a}}{a^3}\right)\nonumber \\
\frac{g_{0 \rho}}{\sqrt{-g}}\frac{\delta S_{GB}}{\delta g_{0 \rho}}&=&\frac{1}{2}(D-1)(D-2)(D-3)(D-4)\frac{\dot{a}^4}{a^4}
\end{eqnarray}
which identically vanishes for $D=4$.
The Einstein tensor $G_{\mu \nu}=R_{\mu \nu}-\frac{1}{2}g_{\mu \nu}R$ has the following non-zero components
\begin{eqnarray}\label{13}
G_{00}&=&\frac{(D-1)(D-2)}{2}\frac{\dot{a}^2}{a^2}\nonumber \\
G_{ii}&=&\frac{(D-2)(3-D)}{2}\dot{a}^2+(2-D)a\ddot{a}
\end{eqnarray}
So collecting all together, by variation of the action (\ref{1}) with respect to the metric, for the time $(00)$ component, we obtain
\begin{equation}\label{14}
\frac{(D-1)(D-2)}{2}M_P^2 H^2=-\alpha(D-4)(D-3)(D-2)(D-1)H^4+\rho_m
\end{equation}
and for $ii$ components we derive
\begin{eqnarray}\label{15}
&&M_P^2\left(\frac{(D-2)(3-D)}{2}H^2+(2-D)\frac{\ddot{a}}{a}\right)=\nonumber \\
&&P_m+\alpha(D-4)(D-3)(D-2)\left((D-5)H^2+4\frac{\ddot{a}}{a}\right)H^2,
\end{eqnarray}
where $H:=\frac{\dot{a}}{a}$ is the Hubble parameter and the matter ingredient in $S_m$ is assumed to be a perfect fluid with energy density $\rho_m$ and pressure $P_m$.
Using the identity $\frac{\ddot{a}}{a}=\dot{H}+H^2$, (\ref{15}) can be rewritten as
\begin{equation}\label{16}
\dot{H}=-\frac{P_m+\rho_m}{(D-2)\left(4\alpha (D-3)(D-4)H^2+M_P^2\right)}
\end{equation}
(\ref{14}) and (\ref{16}) are Friedmann equations in D dimensions. In four dimensions they reduce to the familiar Friedmann equations, unless $(D-4)\alpha$ gains a finite nonzero value. This is only possible for  $\alpha\propto \frac{1}{D-4}$. By setting $\alpha\to \frac{\alpha}{D-4}$, in four dimensions (\ref{14}) and (\ref{16}) reduce to
\begin{eqnarray}\label{17}
3M_P^2H^2&=&\rho_m-6\alpha H^4\nonumber \\
\dot{H}&=&-\frac{P_m+\rho_m}{2\left(4 \alpha H^2+M_P^2\right)}
\end{eqnarray}
The Gauss-Bonnet term modified the Friedmann equations through terms that are functions of $H$ and $\dot{H}$ which in their turns depend on the components that have filled the Universe. So we expect that the modification in the cosmic evolution depends also on $\rho_m$ and its equation of state. The modification of Friedmann equations by higher power of the Hubble parameter occurs also in other models. e.g. see \cite{mod}.

The equations (\ref{17}), imply that the matter satisfies the continuity equation
\begin{equation}\label{28}
\dot{\rho_m}+3H(P_m+\rho_m)=0
\end{equation}

If in an era corresponding to $H_0$, the matter density and the component  $6\alpha H_0^4$ (e.g. as dark energy) have the same order of magnitude, from (\ref{17}) we expect to have $3M_P^2 H_0^2\sim 6\alpha H_0^4$, which results in $\alpha \sim \frac{M_P^2}{2H_0^2}$. Taking $H_0$ as the expansion rate in the present epoch $H_0\sim 10^{-33} eV$ (where the dark energy and matter densities are of the same order), we find  $\alpha \sim 10^{120}$.  This is in agreement with (\ref{ref2}):  If $T^m$ and $\alpha \mathcal{G}$ have the same order of magnitude, we obtain $R\sim \alpha \mathcal{G}/M_P^2$. In four- dimensional FLRW space-time,  $R=6\left(\dot{H}+2H^2\right)$ and $\mathcal{G}=24H^2\left(\dot{H}+H^2\right)$. Therefore $\alpha\sim M_P^2/H^2$ and as in our study the Hubble parameter is much less than the Planck mass (our energy scale is much less than the Planck scale): $H^2\ll M_P^2$,  $\alpha$ must be a large number.  We will investigate this subject also in the next section through the study of the late time acceleration of the Universe, i.e. an epoch in which dark energy becomes relevant.

\section{Cosmic acceleration in four-dimensional Einstein-Gauss-Bonnet cosmology}
Based on modified Friedmann equations, we will study the possible acceleration of FLRW Universe whose dominant matter component is a barotropic matter (e.g. dark matter) $\rho_m$, with pressure $P_m=w_m\rho_m$. In the following, unless we explicitly mention, we assume that the equation of state (EoS) parameter satisfies $w_m\geq -\frac{1}{3}$ (so that it does not act as dark energy or is not responsible for the positive acceleration).

In order for the Hubble parameter to be real, the following condition must hold
\begin{equation}\label{18}
\alpha \rho_m>-\frac{3}{8}M_P^4
\end{equation}
Also the positivity of $\rho_m $ requires
\begin{equation}\label{19}
\alpha H^2\geq -\frac{1}{2}M_P^2
\end{equation}
By substituting $\rho_m$ from the first equation of (\ref{17})in the second one, the deceleration parameter $q=-1-\frac{\dot{H}}{H^2}$ and the EoS parameter of the Universe $w$, are obtained as
\begin{equation}\label{20}
q=-1+\frac{3}{2}\gamma_m\frac{1+2\alpha\frac{H^2}{M_P^2}}{1+4\alpha\frac{H^2}{M_P^2}}
\end{equation}
\begin{equation}\label{200}
w=-1+\gamma_m\frac{1+2\alpha\frac{H^2}{M_P^2}}{1+4\alpha\frac{H^2}{M_P^2}},
\end{equation}
respectively. Where $\gamma_m=w_m+1$. In agreement with our discussion after (\ref{17}), the above equations show that the effect of $S_{GB}$ (in (\ref{1})) on evolution of the Universe depends also on the EoS parameter of $\rho_m$. For example if we took $\gamma_m=0$ (hence $\rho_m$ is a constant), the Gauss-Bonnet contribution would have no effect on the deceleration parameter, i.e. $q=-1$. This can also be verified directly from (\ref{17}), which implies that for $\gamma_m=0$ the Hubble parameter is a constant given by $H^2=\frac{1}{3M_P^2}\rho_\Lambda$, where $\rho_\Lambda$ is an effective cosmological constant determined by $\rho_\Lambda=\frac{-3M_P^4-(+)\sqrt{9M_P^8+24\alpha M_P^4\rho_m}}{4\alpha}$ for $\alpha<(>)0$.

We have acceleration, ($q<0$), provided that
\begin{equation}\label{21}
\gamma_m\frac{1+2\alpha\frac{H^2}{M_P^2}}{1+4\alpha\frac{H^2}{M_P^2}}<\frac{2}{3}
\end{equation}
and for
\begin{equation}\label{22}
\frac{1+2\alpha\frac{H^2}{M_P^2}}{1+4\alpha\frac{H^2}{M_P^2}}<0
\end{equation}
the super-acceleration ($q<-1$ or equivalently $\dot{H}>0$) occurs.

\subsection{\{$\alpha>0,\gamma_m>\frac{2}{3}\}$}
For $\alpha>0$, the acceleration condition (\ref{21}) reduces to
\begin{equation}\label{23}
2\alpha(3\gamma_m-4)\frac{H^2}{M_P^2}<2-3\gamma_m.
\end{equation}
For $\gamma_m>2/3$, acceleration requires $\gamma_m<4/3$. Taking the matter as cold dark matter $\gamma_m=1$, (\ref{23}) becomes
\begin{equation}\label{24}
2\alpha\frac{H^2}{M_P^2}>1
\end{equation}
Hence if one intends to study the present acceleration of the Universe in this context, as $\frac{H_0^2}{M_P^2}\ll 1$ , where $H_0=H(a=1)$ is the present Hubble parameter, he must choose a value of order  $\frac{M_P^2}{H_0^2}$ for $\alpha$. Indeed $\frac{M_P^4}{H_0^2 M_P^2}\sim 10^{120} $ is of the same order as the ratio of the theoretical vacuum energy density to the observed cosmological density, encountered in the cosmological constant problem.  For $H_0=67.4kms^{-1}Mpc^{-1}$ \cite{Planck}, $\alpha$ is obtained as $\alpha=0.4\times10^{120}$.

Note that by considering $\gamma_m>2/3$ and $\alpha>0$, the minimum of the EoS parameter of this Universe is $w_{min.}=-2/3$. For $\gamma_m=1$ (cold dark matter), $w$ cannot be less than $-1/2$. This lies in the range reported in \cite{ecdm}, by considering  an extended cosmological model with respect to the $\Lambda CDM$ and also a global analysis of current cosmological data. For example in case of the 12 parameters model, for the Planck+Lensing it is found $w=-1.03^{+0.62}_{-0.26}$ \cite{ecdm}. Also, If based on $wCDM$ cosmological results \cite{Planck}, one takes the ratio density of dark energy as $\Omega_{d0}=0.68$, and its EoS parameter as $w_d=-0.957\pm 0.08$ (68\%,Planck TT,TE,EE+lowE+lensing), he finds the equation of state parameter of the Universe as $w=w_d\Omega_d+w_m\Omega_m$, which is comparable with the minimum of our model.

\begin{figure}
\centering
\includegraphics[width=5 cm]{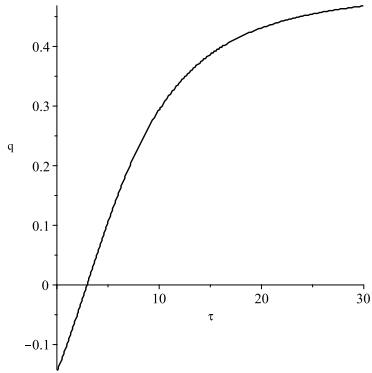}
\caption{Deceleration parameter in terms of dimensionless time $\tau=tH_0$, with $\alpha\frac{H_0^2}{M_P^2}=20$ and initial conditions ${\frac{\rho_m(0)}{M_P^2H_0^2}}=0.9$}
\label{fig1}
\end{figure}

Although it seems that there exists acceleration for this model, but what about the transition from deceleration to acceleration?
For an expanding Universe in order that these transitions occur, we require to have $\frac{dq}{dH^2}>0$ at the transition point. From (\ref{20}) we have
\begin{equation}\label{25}
\frac{dq}{dH^2}= -\frac{3\alpha\gamma_m M_P^2}{(M_P^2+4\alpha H^2)^2}
\end{equation}
which is negative for $\alpha>0$. So in the $\alpha>0$ case, although an acceleration solution may exist, the transition cannot be described by this special model alone. As an illustrative example, By using (\ref{17}) and (\ref{28}),  for $\gamma_m=1$, we have depicted the deceleration parameter in the case of  $\alpha>0$ in fig.(\ref{fig1}), in terms of dimensionless time $\tau=t H_0$, where $H_0=H(a=1)$, showing an accelerated expanding Universe which ends to a deceleration state. At $a=1$, we have taken ${\frac{\rho_m(0)}{M_P^2H_0^2}}=0.9$ \cite{Planck}.

Based on (\ref{24})one can argue that:  if today the Universe is in an accelerated phase, it has also been in this phase in the past, provided that $\gamma_m=1$. Hence in this situation, we have acceleration from the beginning of the matter-dominated era (in the sense that when $\rho\simeq \rho_m$) preventing structure formation. Although the model does not support deceleration to an acceleration phase transition during one era, this transition may happen when the Universe goes from one era to another one, e.g. from (\ref{23}) it is clear that the Universe had a deceleration phase in the radiation dominated era ($\gamma_m=4/3$). Therefore the transition from deceleration to acceleration happens between radiation dominated and matter-dominated era $\rho_{radiation}\sim \rho_{matter}$. But in many cosmological models, the present acceleration began during the matter-dominated era. By this assumption,  it seems the model may be only valid for the late time after the transition. In this situation, to extend the study to a larger time, e.g. one may consider an extended model in which $\alpha$ appears as a function of a dynamical field which becomes active in the matter-dominated era such that before that era the coefficient of the Gauss-Bonnet term vanishes and leaves us with a decelerating non modified standard model. Some examples of such a model, where dark energy becomes active in the matter-dominated era and gives rise to the onset of dark energy, in the context of the screening and the modified gravity, can be found in \cite{sym0,sym1,sym2,sym3,sym4}.

\subsection{$\{\gamma_m>\frac{2}{3} ,\alpha<0\}$}

Taking $\alpha<0$, we find a singularity at $4\alpha\frac{H^2}{M_P^2}=-1$, diving the problem into two branches $4\alpha\frac{H^2}{M_P^2}>-1$ and $4\alpha\frac{H^2}{M_P^2}<-1$.

For $4\alpha\frac{H^2}{M_P^2}<-1$ and by considering (\ref{19}), from (\ref{22}) we find that the Universe is in a super-accelerated regime. In this case $|\alpha|>\frac{1}{4}\frac{M_P^2}{H^2}$.  So again if one tries to attribute the positive acceleration to the Gauss-Bonnet term, far away from the quantum gravity regime, he must choose a large value for $|\alpha|$.

For $4\alpha\frac{H^2}{M_P^2}>-1$ and by considering (\ref{19}), we obtain
\begin{equation}\label{26}
\frac{1+2\alpha\frac{H^2}{M_P^2}}{1+4\alpha\frac{H^2}{M_P^2}}>1
\end{equation}
So by comparing with (\ref{21}), we find out that the Universe is in decelerated regime.

For $\alpha<0$,
\begin{equation}\label{27}
\frac{dq}{dH^2}= -\frac{3\alpha\gamma_m M_P^2}{(M_P^2+4\alpha H^2)^2}>0
\end{equation}
Therefore $q$ increases in the super-acceleration regime, and tends to $q=-1$ eventually, while it decreases in the deceleration regime and tends to $q=-1+\frac{3}{2}\gamma_m$, without crossing it (assuming that $\rho_m$ is still the only relevant matter component). So like the $\alpha>0$ case, we are unable to describe transitions in this context. As we will see later, $q=-1$  is a stable fixed point.

\begin{figure}[H]
\centering
\includegraphics[width=5 cm]{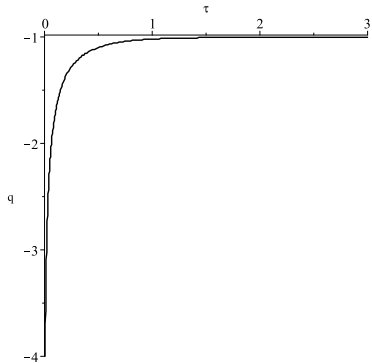}
\caption{Deceleration parameter in terms of dimensionless time $\tau=tH_0$, with $\alpha\frac{H_0^2}{M_P^2}=-0.4$ and initial conditions${\frac{\rho_m(0)}{M_P^2H_0^2}}=0.9$}
\label{fig2}
\end{figure}

\begin{figure}[H]
\centering
\includegraphics[width=5 cm]{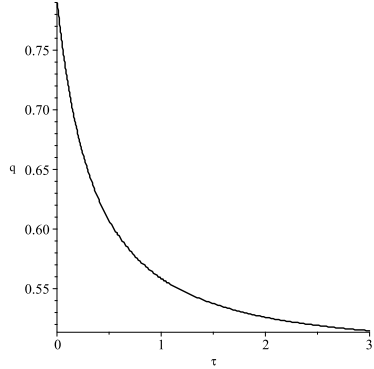}
\caption{Deceleration parameter in terms of dimensionless time $\tau=tH_0$, with $\alpha\frac{H_0^2}{M_P^2}=-0.2$ and initial conditions ${\frac{\rho_m(0)}{M_P^2H_0^2}}=0.9$}
\label{fig3}
\end{figure}

By using (\ref{17}) and (\ref{28}), we have depicted the deceleration parameter in the case of $\alpha<0$ for $4\alpha\frac{H^2}{M_P^2}<-1$ and $4\alpha\frac{H^2}{M_P^2}>-1$, in terms of dimensionless time $\tau=tH_0$, where $H_0=H(a=1)$, showing a super-acceleration expanding Universe which tends asymptotically to the fixed point $q=-1$ in the former (see fig.(\ref{fig2})), and a decelerating Universe tending to $q=1/2$ eventually in the latter for $\gamma_m=1$ (see fig.(\ref{fig3})) .

In the absence of any matter $\rho_m=0$, and for $\alpha<0$, we have de Sitter solution characterized by the constant Hubble parameter $H^2=-\frac{M_P^2}{2\alpha}$. This is a critical point, and is stable.  (\ref{17}) and ( \ref{28}) form a system of autonomous equations with the critical point $\{\bar{H}^2=-\frac{M_P^2}{2\alpha}, \bar{\rho}_m=0\}$. Perturbing the system around its critical point $H=\bar{H}+\delta H,\,\, \rho_m=\bar{\rho}_m+\delta\rho$, we find
\begin{equation}\label{29}
\delta \dot{\rho_m}+3\bar{H}\gamma_m \delta\rho_m=0
\end{equation}
and
\begin{equation}\label{30}
-M_P^2\left(1+\frac{4\alpha}{M_P^2}\bar{H}^2\right)\delta\dot{H}=\frac{1}{2}\gamma_m\delta\rho_m
\end{equation}
After some computation, we derive
\begin{equation}\label{31}
\delta \rho_m=-6M_P^2\bar{H}\delta H
\end{equation}
Therefore from  (\ref{30}), (\ref{31}), we obtain
\begin{eqnarray}\label{32}
\delta \dot{H}=-3\gamma_m\bar{H} \delta H
\end{eqnarray}
Hence
\begin{equation}\label{33}
\delta \rho_m\propto \delta H\propto e^{-3\gamma_m \bar{H}t}
\end{equation}
So for an expanding Universe $\bar H>0$, the critical point is stable.

\subsection{Gauss-Bonnet term as a dark energy component}

For $\alpha<0$, one may attribute an energy density $\rho_d=-6\alpha H^4$, and a pressure $P_d$ to the Gauss-Bonnet contribution, and rewrites the Friedmann equations as
\begin{eqnarray}\label{34}
&&H^2=\frac{1}{3M_P^2}(\rho_m+\rho_d)\nonumber \\
&&\dot{H}=-\frac{1}{2M_P^2}(\gamma_m \rho_m +P_d+\rho_d)
\end{eqnarray}
By comparing (\ref{34}) with (\ref{17}), after some calculations we derive
\begin{equation}\label{35}
P_d=\frac{2\alpha H^2(12\alpha H^4+3M_P^2 H^2-2\gamma_m \rho_m)}{4\alpha H^2+M_P^2}
\end{equation}

There is a fundamental difference between $\rho_d=-6\alpha H^4$ and other energy densities like $\rho_m$. $\rho_d$, and $P_d$ act effectively as energy density and pressure in the Friedmann equations and, through their definitions, depend on the Hubble parameter and other components filling the Universe. So $w_d$ is not expected to be an independent quantity. We can attribute an energy-momentum tensor to this effective component whose the trace is obtained from (\ref{ref2}) as $T^{(d)}=\alpha \mathcal{G}$, giving $3P_d-\rho_d=24\alpha H^2(\dot{H}+H^2)$. $w_d$ is derived by dividing (\ref{35}) by $-6\alpha H^4$:
\begin{equation}\label{36}
w_d=-1+2\frac{\gamma_m\Omega_m}{1+4\alpha \frac{H^2}{M_P^2}}
\end{equation}
Hence, this EoS parameter depends completely on the matter's EoS parameter and its ratio density $\Omega_m:=\frac{\rho_m}{3M_P^2H^2}$. For example, in the absence of matter $\Omega_m=0$ or for $\gamma_m=0$, we obtain $w_d=-1$ and the Gauss-Bonnet term behaves as a cosmological constant. This can also be derived from (\ref{17}), which states that whatever the $\alpha$ value, the Hubble parameter is a constant for $P_m+\rho_m=0$.

The ratio density of the dark sector is $\Omega_d=\frac{\rho_d}{3M_P^2H^2}$, but $\rho_d=-6\alpha H^4$, therefore
\begin{equation}\label{37}
\Omega_d=-2\alpha\frac{H^2}{M_P^2}
\end{equation}
This equation may be employed to obtain $\alpha$. If at $H=H_0$, $\Omega_d=\Omega_{d0}\sim \Omega_{m0}$, then $\Omega_{d0}\sim 1$. Hence from (\ref{37}) we find: $\alpha\sim  \frac{M_P^2}{2H_0^2}$, which is in agreement with our discussion in the last paragraph of the previous section.

Putting $\alpha=-\frac{1}{2}\Omega_d\frac{M_P^2}{H^2}$ back into (\ref{36}) gives the EoS parameter as
\begin{equation}\label{38}
w_d=-1+2\gamma_m\frac{1-\Omega_d}{1-2\Omega_d}
\end{equation}
From $w=w_m\Omega_m+w_d\Omega_d$, the EoS of the Universe is derived as
\begin{equation}\label{39}
w=\frac{(\gamma_m-2)\Omega_d+1-\gamma_m}{2\Omega_d-1}
\end{equation}

For $\Omega_d<\frac{1}{2}$, we have $w_d>1/3$ and $w>-1/3$ and the Gauss-Bonnet term plays the role of a matter with positive pressure. As a result, in our study where $w_m\geq {-1/3}$,  ${\Omega_d<1/2}$ corresponds to a decelerating phase. In this situation as $H$ is decreasing, the system eventually tends to $\{H=0, \rho_m=0, q=0.5\}$.

For ${\Omega_d>1/2}$, we have $w<-1$ and $w_d<-1$, and the Gauss-Bonnet term plays the role of a phantom-like component which describes a super-acceleration phase. If the matter is considered as a pressureless dark matter then $w_d=\frac{1}{1-2\Omega_d}$ and $w=\frac{\Omega_d}{1-2\Omega_d}$. In this phase $\frac{dq}{dH^2}>0$ and as $H$ increases, $q$ increases too, and the system tends to $w=-1$  which is a stable fixed point as proved before. In this situation too, we have only a super-acceleration regime, and the model is unable to describe the possible transition to or from (normal)acceleration regime.

The behavior of the Hubble parameter in terms of the redshift can be derived from (\ref{17}) which can be rewritten as
\begin{equation}
\frac{H^2}{H_0^2}=\Omega_{m0}(1+z)^3-\frac{6\alpha H^4}{3M_P^2H_0^2}
\end{equation}
where the subscript "0" denotes the value at $a=1$ or $z=0$. This may be rewritten in terms of ratio densities as
\begin{equation}
\frac{\Omega_d}{\Omega_{d0}}-\frac{\Omega_d^2}{\Omega_{d0}}=\Omega_{m0}(1+z)^3
\end{equation}
At $z=-1$ we have $\Omega_{d}=1$, which as stated before shows the eventual fate of the system.  As in the acceleration phase $\Omega_d>\frac{1}{2}$, we derive the range of the validity of the model as
\begin{equation}\label{r10}
z<(4\Omega_{d0}\Omega_{m0})^{-\frac{1}{3}}-1
\end{equation}
Note that in $\Lambda CDM$ model, $q<-1$ is excluded.  Recently, within the study of $H_0$ tension, discussions about phantom like dark energy and the possibility to have $q<-1$ raise again.  E.g. in \cite{Marra},  where local determination of the Hubble constant and the deceleration parameter is studied,
the deceleration parameter at $z=0$ is constrained to  $q_0=-1.08\pm 0.29$. Taking $q_0=-1.37$, we obtain $\Omega_{d0}\simeq 0.8$, and (\ref{r10}) becomes $z<0.2$. From (\ref{r10}), we find that the model may describe an acceleration only in a restricted period in low redshift, and the Gauss-Bonnet modification  behaves effectively as a phantom like dark energy emerged in the late time. In this situation too, as discussed in the paragraph before the subsection 3.2, to extend the study to the deceleration epoch, an extended model which reduces to the present model in the late time is required.

\section{Conclusion}
We considered the model introduced in \cite{GBN}, where the Gauss-Bonnet term appears with a singular coefficient in the action. This singularity is eliminated in the equations of motion in four dimensions, and new contributions from the Gauss-Bonnet term emerge, giving rise to physical results. In an FLRW space-time filled nearly with a barotropic matter, we precisely derived the modified Friedmann equations and studied generally the cosmological consequences of the new terms in the acceleration of the Universe. To avoid the influence of other dark energies in our results, we restricted our model to contain only matter whose equation of state parameter satisfies $w_m>-1/3$. Based on Friedmann equations, the conditions required to have accelerating solutions were derived. It was shown that when the coefficient sign is positive we may obtain a solution with positive acceleration but the deceleration to acceleration transition cannot be explained by this model during a pressureless matter-dominated era and eventually, the Universe decelerates.

For the negative coefficient, the Gauss-Bonnet term imitates the role of a (dark energy) component whose equation of state parameter is a function of its ratio density $\Omega_d$ and the equation of state parameter of the other ingredient($w_m$). The solutions were classified into two distinct sets $\Omega_d<1/2$ and $\Omega_d>1/2$ separated by a singularity at $\Omega_d=1/2$. For $\Omega_d>1/2$, the Universe is in a super-accelerated phase in a low redshift and eventually tends to a de Sitter stable fixed point at $z=-1$. The deceleration parameter lies in the domain reported in the literature in the study of $H_0$ tension in the context of the dynamical phantom dark energy model. For $\Omega_d<1/2$, the Gauss-Bonnet term acts as a matter with positive pressure and the Universe is in the deceleration phase.

In our study, we found that to obtain a cosmic acceleration in the present epoch, the regularized coefficient of the Gauss-Bonnet term must have the same order as the discrepancy of vacuum and observable dark energy densities encountered in the cosmological constant problem. This lies in the term playing the r\^{o}le of dark energy in this model. To elucidate this, we remind that at the present epoch the dark matter and dark energy densities have the same order of magnitude $\rho_{m0}\sim \rho_{d0}$. By using the Friedmann equation (\ref{17}), we find that $\rho_{m0}\sim 3M_P^2H_0^2$. Therefore  $\rho_{d0}\sim \rho_{m0}\sim 3M_P^2H_0^2$. But the dark energy contribution is coming from the Gauss-Bonnet invariant which behaves as $\sim H_0^4$ and for $\alpha<0$ is related to an effective density $\rho_d=-6\alpha H_0^4$. Hence $|\alpha| H_0^4\sim M_P^2 H_0^2$, leading to $|\alpha| \sim \frac{M_P^2}{H_0^2}$. Therefore if the theory is valid in regions far from quantum gravity (i.e. $H_0^2\ll M_P^2$) as is our epoch, we expect to obtain a huge value for
$|\alpha|$.

In all situations, it seems that to extend the study to the deceleration matter-dominated phase before the acceleration, we need to extend the model such that the extended model reduces to the actual model in the late time.  As an outlook this may be done by considering a dynamical coefficient for the Gauss-Bonnet invariant such that it becomes active only in the low redshift, playing the role of the dark energy. E.g. similar to the screening models, one can consider $\alpha=\alpha(\phi)$ where $\phi$ is a scalar field such that at higher redshift $<\alpha(\phi)>=0$ and for lower redshift $<\alpha(\phi)>\neq 0$.  Adding new degrees of freedom to the model has also been used as a regularization technique \cite{Man}. This technique consists of considering an extra scalar degree of freedom, $\phi(x)$, inserted through a conformal transformation $\tilde{g}_{\mu \nu}=e^{2\phi}g_{\mu \nu}$. One may regularize the theory for $D\to 4$, by adding the counter term $\frac{\alpha}{D-4}\int d^4x \sqrt{-\tilde{g}}\tilde{\mathcal{G}}$, and expanding around $D=4$, to obtain a divergence-free action leading to covariant well defined field equations comprising non-trivial contributions from the Gauss-Bonnet term\cite{clif}. Although the field equations derived from this method contain also the scalar field and its derivatives, by taking the trace of the field equation, one obtains the same equation as (\ref{ref2})\cite{clif}. Here too in order that $\frac{\alpha}{2}\mathcal{G}$ has the same order of magnitude as the matter sector, we must have  $\alpha\sim \frac{M_P^2}{H_0^2}$.

\end{document}